\documentclass[aps,pra,twocolumn,amsmath,amssymb,superscriptaddress,floatfix]{revtex4-2}

\usepackage[utf8]{inputenc}
\usepackage[T1]{fontenc}
\usepackage{microtype}
\usepackage{amsmath}
\usepackage{graphicx}
\usepackage{times}
\usepackage{amssymb}
\usepackage{mathtools}
\usepackage{txfonts}
\usepackage{physics}
\usepackage{float}
\usepackage[table,xcdraw]{xcolor}
\usepackage{mathrsfs}
\usepackage[colorlinks = true, linkcolor = blue, urlcolor  = blue, citecolor = blue, anchorcolor = blue]{hyperref}
\usepackage{tikz}
\usepackage{bm}
\usepackage{tabularx}
\usepackage{pgfplots}
\usepackage[version=4]{mhchem}
\usepackage[caption = false, subrefformat=parens,labelformat=parens]{subfig}
\usepackage{ragged2e} % for the \justifying macro

\makeatletter

\def\CT@@do@color{%
  \global\let\CT@do@color\relax
        \@tempdima\wd\z@
        \advance\@tempdima\@tempdimb
        \advance\@tempdima\@tempdimc
\advance\@tempdimb\tabcolsep
\advance\@tempdimc\tabcolsep
\advance\@tempdima2\tabcolsep
        \kern-\@tempdimb
        \leaders\vrule
                \hskip\@tempdima\@plus  1fill
        \kern-\@tempdimc
        \hskip-\wd\z@ \@plus -1fill }
\makeatother

\pgfplotsset{compat=newest}
\DeclareCaptionJustification{justified}{\justifying}
\newcommand{\DetParam}{\Delta_\text{Det}}
\newcommand{\Phase}{\Phi}
\newcommand{\Grav}{\phi}
\newcolumntype{Y}{>{\centering\arraybackslash}X}
\definecolor{mygreen}{RGB}{1,113,0}

\begin{document}
\title{Local Measurement Scheme of Gravitational Curvature using Atom Interferometers}

\author{Michael Werner$^\ast$}
\affiliation{Institut für Theoretische Physik, Leibniz Universität Hannover, Appelstraße 2, 30167 Hannover, Germany}

\author{Ali Lezeik}
\affiliation{Institut für Quantenoptik, Leibniz Universität Hannover, Welfengarten 1, 30167 Hannover, Germany}

\author{Dennis Schlippert}
\affiliation{Institut für Quantenoptik, Leibniz Universität Hannover, Welfengarten 1, 30167 Hannover, Germany}

\author{Ernst M. Rasel}
\affiliation{Institut für Quantenoptik, Leibniz Universität Hannover, Welfengarten 1, 30167 Hannover, Germany}

\author{Naceur Gaaloul}
\affiliation{Institut für Quantenoptik, Leibniz Universität Hannover, Welfengarten 1, 30167 Hannover, Germany}

\author{Klemens Hammerer}
\affiliation{Institut für Theoretische Physik, Leibniz Universität Hannover, Appelstraße 2, 30167 Hannover, Germany}

\date\today

\begin{abstract}
Light pulse atom interferometers (AIFs) are exquisite quantum probes of spatial inhomogeneity and gravitational curvature. Moreover, detailed measurement and calibration are necessary prerequisites for very-long-baseline atom interferometry (VLBAI). Here we present a method in which the differential signal of two co-located interferometers singles out a phase shift proportional to the curvature of the gravitational potential. The scale factor depends only on well controlled quantities, namely the photon wave number, the interferometer time and the atomic recoil, which allows the curvature to be accurately inferred from a measured phase. As a case study, we numerically simulate such a co-located gradiometric interferometer in the context of the Hannover VLBAI facility and prove the robustness of the phase shift in gravitational fields with complex spatial dependence. We define an estimator of the gravitational curvature for non-trivial gravitational fields and calculate the trade-off between signal strength and estimation accuracy with regard to spatial resolution. As a perspective, we discuss the case of a time-dependent gravitational field and corresponding measurement strategies.
\end{abstract}

\maketitle

\section{Introduction}

AIFs are high-precision instruments used in a wide variety of research fields. Their versatility includes tasks such as determining the fundamental constants~\cite{parker2018measurement, morel2020determination,rosi2014precision, schelfhout2024singlephoton}, serving as quantum sensors to measure Earth's gravitational field~\cite{Wu_2019, del_Aguila_2018, Fang_2016}, proposing measurements for gravitational wave detection~\cite{beaufils2022coldatom, chen2023enhancing, canuel2020elgar}, exploring fundamental physics and alternative gravitational models~\cite{schlippert2014quantum, damour2012theoretical, colladay1997cpt,Asenbaum2020}, and performing measurements related to time dilation and gravitational redshift~\cite{loriani2019interference, roura2021measuring, zych2011quantum, diPumpo2021GravitationalRedshift}. In particular, their accuracy as sensors of gravitational fields and their gradients is becoming increasingly important for applications in civil engineering~\cite{Stray2022}, inertial sensing~\cite{Geiger2020b} and geodesy~\cite{Antoni-Micollier2022,Bidel2018,Bidel2020,Farah2014,Haagmans2020,Leveque2024}.

AIFs are utilized to measure the gravitational field, there they provide information about the linear gravitational acceleration $g$ along the atomic trajectory. This approach is highly accurate because the leading order phase shift $\Delta \Phi = g k T_R^2$ connects the desired value of $g$ with the wave vector $k$ and the interferometer time $T_R$, both of which are known with very high precision. 
For measuring the (constant) gravitational gradient, a gradiometric experimental setup is employed, involving a comparison of $g$-measurements from two spatially separated gravimeters, effectively interpolating the $g$ values between their spatial positions. Such gradiometric experiments are theoretically limited by the measurement uncertainty of the phase shift and the uncertainty of the height difference between the two interferometers. Another way to extract knowledge about the gravity gradient is done using more elaborate AIF geometries~\cite{Figure8}. In these cases, however, the phase shift depends non-linearly on the gravitational field, making an estimation more complicated.

State-of-the-art AIFs are being constructed with increasingly longer baselines \cite{coleman2018magis100, badurina2020aion, zhan2020zaiga, abend2024terrestrial} and more efficient large momentum transfer (LMT) techniques \cite{PhysRevLett.121.133201, rodzinka2024optimal, li2024sensitivity}, extending beyond the region where the assumption of a constant gradient of the gravitational field remains valid. 
The transition to non-trivial gravitational curvature is not only a challenge for large baseline interferometers, but can also be seen as an opportunity for experiments with gravitational test masses. Deliberately introduced non-trivial gravitational fields, which allow the measurement of phases along the atomic trajectory to probe this non-linearity, have been exploited in~\cite{asenbaum2017phase, asenbaum2024matter} and led to the proposed gravitational Aharonov-Bohm effect~\cite{overstreet2022observation}. Measuring anomalies in the gravitational gradient is also used to detect inhomogeneities in the gravitational field~\cite{Stray2022} and will become evermore important for civil engineering and quantum metrology. Resolving a spatially varying gravity gradient to high accuracy with a gradiometric AIF setup is, however, equivalent to comparing $g$-measurements in close proximity. This procedure is therefore increasingly error prone, because of the relative uncertainty in the position of the atomic ensembles, compared to the separation of the two constituent AIFs.

\begin{figure}
    \centering
    \includegraphics[width = \linewidth]{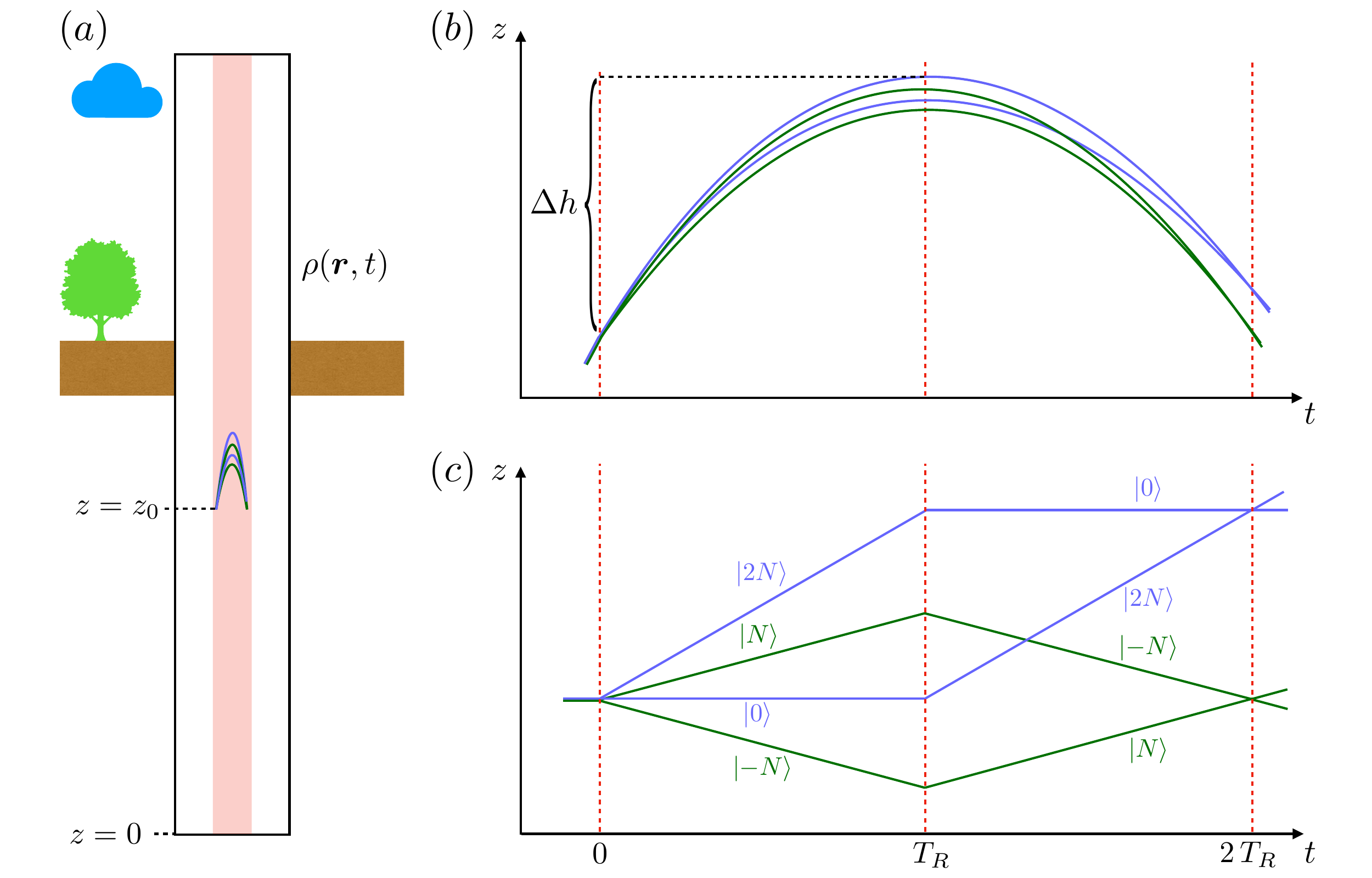}
    \caption{Depiction of the co-located gradiometric interferometer (CGI) setup consisting of a SDDI (green) and a MZI (blue) in a gravitational field sourced by the mass density $\rho(\bm{r}, t)$. (a) Position of the CGI in a large baseline interferometry setup as as determined by the initial height $z_0$. CGI geometry shown in more detail (b) in the laboratory frame and (c) in the freely falling frame. $\ket{N}$ denotes a momentum eigenstate with $N$ momentum quanta, as compared to the initial wave packet. The speed of light was set infinite for the laser pulses in this plot.}
    \label{Fig: Tidal_Phase_Geometry}
\end{figure}

In this analysis, we introduce a novel geometry for AIFs that is exclusively sensitive to the gravitational curvature, that is, the gradient and higher-order derivatives of the gravitational field. The resulting phase depends, in an idealized model, on the gravity gradient, $k$, $T_R$ and, additionally, on the atomic recoil $\hbar/m$, which is also known to high precision. Notably, such an AIF geometry does not require two distinct and spatially separated experimental setups; instead, it consists of two co-located AIFs, initialized at the same height. As a result, the measurement resolution of such a co-located gradiometric interferometer (CGI) is solely determined by the signal magnitude and is not constrained by a spatial separation between the constituent AIFs. 

As a case study, we simulate CGI schemes for the VLBAI facility in Hannover, using its precisely known gravitational field \cite{schilling2020vertical, lezeik2023understanding}, and analyze the trade-off between signal strength and spatial resolution. Since the atoms sample and average a macroscopic portion of the gravitational field during their flight time, it is not clear a priori how to infer the gravitational curvature at a particular height from a measured phase shift. Our analysis will address and resolve this issue for the CGI and define a general estimator of the gravitational curvature. We also highlight the importance of achieving temporal resolution of the gravitational field, and discuss how this novel AIF geometry might help to accomplish this task.

\section{Results}

\subsection{Measurement of gravitational curvature}\label{Section_2}

Throughout this analysis, we will assume one-dimensional movement of the AIF atoms along the z-axis of a local coordinate system, originating at a fixed height of the experiment and disregard Earth's rotation. We denote the gravitational potential in the vicinity of the experimental setup by $\Grav(z)$. Expanding this potential in a neighborhood of the origin of this local coordinate system yields 
\begin{align}
    \Grav(z) &= \phi_0 + g \, z + \sum_{n=2} \frac{\Grav^{(n)}}{n!} z^n &
    \Gamma(z) = \pdv[2]{\Grav(z)}{z},
\end{align}
where the summation must be carried to a considerable order, depending on the complexity of the gravitational environment, as we will discuss below. To be specific we refer to all the terms in $\Gamma(z)$ as gravitational curvature, i.e. the gravity gradient and higher order derivatives of the potential. Note that curvature, in a general relativistic sense, is defined via components of the Riemann curvature tensor. Those components are -- to first order -- second derivatives of the gravitational potential, as it is the case here \footnote{Note that geodesists refer to gravitational curvature as third order derivatives of the gravitational potential, i.e. the spatial variation of the gravity gradient.}. We will focus on the CGI depicted in Fig.~\ref{Fig: Tidal_Phase_Geometry}, consisting of a MZI with $2 N \hbar k$ momentum transfer and a `Symmetric Double Diffraction Interferometer' (SDDI) with an initial photon kick of $N \hbar k$ in each direction. Regarding potential experimental implementations of the first beam splitter pulse of this geometry, we refer to the analysis in Appendix~\ref{Appendix_C}.

The key finding of the work is, that the differential phase shift of the CGI geometry is dominantly given by
\begin{subequations}
\begin{align}
    \Delta \Phase_\text{MZI} - \Delta \Phase_\text{SDDI} \approx \Delta \Phase_\text{Curv},
    \label{eq: Phase_Approximation}
\end{align}
where $\Delta \Phase_\text{Curv}$ is related to gravitational curvature 
\begin{align}
    \Delta \Phase_\text{Curv} = - \frac{m}{\hbar} \sum\limits_{n=2} \frac{ \Grav^{(n)}}{n!} \qty[\mathcal{A}_\text{MZI}(n) - \mathcal{A}_\text{SDDI}(n)],
    \label{eq: Tidal Phase Abstract Definition}
\end{align}
and $\mathcal{A}_\text{MZI}(n)$, $\mathcal{A}_\text{SDDI}(n)$ are geometry dependent quantities. Denoting the classical solutions of the atomic trajectories on the upper and lower AIF path of the MZI as $z_\text{up/low}^\text{MZI}(t)$ respectively, one can write $\mathcal{A}_\text{MZI}(n)$ as
\begin{align}
    \mathcal{A}_\text{MZI}(n) = \int\limits_0^{2 T_R} \qty(z_\text{up}^\text{MZI} (t)^n - z_\text{low}^\text{MZI} (t)^n) \, \dd t,
\end{align} 
\end{subequations}
which coincides with the spacetime area of the MZI for $n=1$. The formula for the SDDI is completely analogous. Approximations underlying Eq.~\eqref{eq: Phase_Approximation} are discussed in Appendix~\ref{Appendix_A}. Additional phase contributions arising from finite speed of light (FSL) and their mitigation are discussed in Appendix~\ref{Appendix_B}. 
Note that there is no $n = 1$ contribution in $\Delta \Phase_\text{Curv}$. This means that  phase shifts resulting from linear gravitational acceleration $g$ cancel in this geometry to leading order, as we will demonstrate on the example of an idealized gravitational potential of second order in Sec.~\ref{Section_3} and for the concrete  gravitational field of VLBAI Hannover in Sec.~\ref{Section_4}.

In order to analyze phases and their origins, we recapitulate that the phase difference at the output port of an AIF originates from three main components~\cite{hogan2008light, borde1989atomic, storey1994feynman, Kasevich_Chu}: the propagation phase $\Delta \Phase_\text{Prop}$, the kick phase $\Delta \Phase_\text{Kick}$, and the separation phase $\Delta \Phase_\text{Sep}$. The propagation phase is calculated as the difference of the action functional along the upper and lower atomic path. If we denote the classical Lagrangian of an atom of mass $m$ as $L(z(t)) = \frac{m}{2}\dot{z}(t)^2 - m \Grav \qty(z(t))$, the propagation phase can be compactly written as
\begin{align}
    \Delta \Phase_\text{Prop} = - \frac{1}{\hbar} \int\limits_{T_\text{initial}}^{T_\text{final}} L(z_\text{up}(t)) - L(z_\text{low}(t)) \, \dd t,
    \label{eq: Definition_Propagation_phase}
\end{align}
where $z_\text{up}(t)$ and $z_\text{low}(t)$ are the solutions to the classical equations of motion for the atoms on the upper and lower trajectory in the time interval $\qty[T_\text{initial}, T_\text{final}]$, and the global minus sign is conventional. We note that if the gravitational potential is of at most quadratic order in $z$, the expression for the propagation phase is exact and not merely an approximation.

The kick phase is determined by the difference in the imprinted AIF laser phases along each path. It contributes positively to the total phase when a photon is absorbed and negatively when a photon is emitted into the interferometry light fields. Finally, the separation phase is computed by multiplying the average output momentum at the output port with the separation of the atomic wave packets. Both the kick and separation phases are of lesser importance for the current analysis, as the propagation phase will encompass all relevant effects. Furthermore, the separation phase is usually compensated, as we will discuss below.

\subsection{Idealized gravitational fields}\label{Section_3}

The majority of phase contributions in an AIF scale with the enclosed spacetime area, which is identical in both the MZI and SDDI configurations shown in Fig.~\ref{Fig: Tidal_Phase_Geometry}. However, $\Delta \Phase_\text{Curv}$ arises exclusively in the MZI and vanishes in the SDDI, indicating that this particular phase contribution remains unchanged in the differential setup. To gain a deeper understanding, we will now analyze the reasons for this behavior within the context of an idealized gravitational potential
\begin{align}
    \Grav_\text{Ideal}(z) = g z + \frac{1}{2} \Gamma_0 z^2, \quad \text{i.e.,} \quad \Gamma(z) = \Gamma_0 , \label{eq: Ideal_Gravitational_Potential}
\end{align}
consisting of linear acceleration $g$ and a constant gravity gradient $\Gamma_0$. Solving the Euler-Lagrange equation for this potential with initial conditions $z_0$ and $v_0$ results in an atomic trajectory of the form
\begin{align}
    z(t) = z_0 + v_0 t  - \frac{1}{2} g t^2 
    - \frac{1}{2} \Gamma_0 \qty(z_0 t^2 + \frac{1}{3} v_0 t^3 - \frac{1}{12} g t^4).
\end{align}
When evaluating the the trajectories for the upper and lower AIF paths of the MZI and SDDI, cf. Fig.~\ref{Fig: Tidal_Phase_Geometry}, we assume identical initial conditions $z_0$ and $v_0$ for both interferometers before the first beam splitter, and account for the different photon recoils by the respective momentum kicks at $t=0$. Evaluating the trajectories for the AIF paths at the time $t = T_R$ of the mirror pulse results in positions 
\begin{subequations}
\begin{align}
    z_\text{up}^\text{MZI} \qty(T_R) &= z(T_R) + 2 N \frac{\hbar k T_R}{m}, \\
    z_\text{low}^\text{MZI} \qty(T_R) &= z(T_R) ,
\end{align} \label{eq: Initial_Heights_Second_IF_Interval_MZI}%
\end{subequations}
for the upper and the lower arm of the MZI, and analogously for the SDDI in
\begin{subequations}
\begin{align}
    z_\text{up}^\text{SDDI} \qty(T_R) &= z(T_R) + N \frac{\hbar k T_R}{m}  , \\
    z_\text{low}^\text{SDDI} \qty(T_R) &= z(T_R) - N \frac{\hbar k T_R}{m}.
\end{align}
\label{eq: Initial_Heights_Second_IF_Interval_SDDI}
\end{subequations}
These relations hold up to corrections of order $\mathcal{O}(\Gamma_0)$ which contribute to the resulting phase in negligible order $\mathcal{O}(\Gamma_0^2)$. It is the asymmetry of the trajectories in Eqs.~\eqref{eq: Initial_Heights_Second_IF_Interval_MZI} and \eqref{eq: Initial_Heights_Second_IF_Interval_SDDI} due to the photon recoil that ultimately leads to differences in the sensitivity of the MZI and SDDI regarding gravitational curvature. 

This asymmetry in particular affects the propagation phases, which are
\begin{subequations}\label{eq: Propagation_phase_difference_statement}
\begin{align}
    \Delta \Phase_\text{Prop}^\text{SDDI} &= - \frac{m}{\hbar} \int\limits_{0}^{2 T_R} \qty[ L\qty(z_\text{up}^\text{SDDI}(t)) - L\qty(z_\text{low}^\text{SDDI}(t)) ] \, \dd t, \\
    \Delta \Phase_\text{Prop}^\text{MZI} &\approx \frac{m \Gamma_0}{2 \hbar} \int\limits_{T_R}^{2 T_R} \qty(2 N \frac{\hbar k T_R}{m} )^2   \, \dd t + \Delta \Phase_\text{Prop}^\text{SDDI}\label{eq: MZI},
\end{align}
\end{subequations}
where Eq.~\eqref{eq: MZI} holds up to negligible relativistic corrections (commented on in Appendix~\ref{Appendix_A}). The additional phase in the MZI results from the propagation phase in the time interval $[T_R, 2 T_R]$ along the $\frac{m}{2} \Gamma_0 z(t)^2$ part of the Lagrangian -- especially due to a non-vanishing photon-recoil asymmetry in the initial heights in $z(t)$, as seen in Eqs.~\eqref{eq: Initial_Heights_Second_IF_Interval_MZI} and \eqref{eq: Initial_Heights_Second_IF_Interval_SDDI}. In contrast, this contribution vanishes in the SDDI, since both $z_\text{up}^\text{SDDI} \qty(T_R)^2$ and $z_\text{low}^\text{SDDI} \qty(T_R)^2$ will exhibit identical photon-recoil dependent phase contributions, thereby nullifying any output signal. In a differential measurement setup one is therefore left with a net contribution of propagation phases consisting of the phase of interest
\begin{align}
    \Delta \Phase_\text{Prop}^\text{MZI} - \Delta \Phase_\text{Prop}^\text{SDDI} \approx \frac{2 \Gamma_0 N^2 \hbar k^2 T_R^3}{m} = f \cdot \Gamma_0.
    \label{eq: Tidal_Phase_Ideal_Potential}
\end{align}
Here we introduced a scale factor $f = 2 N^2 \hbar k^2 T_R^3/m$, which translates the phase shift into a value of the (idealized) gravitational gradient $\Gamma_0$. A complete evaluation of phases as outlined before reveals that this is in fact the dominant signal in the differential phase of a MZI and a SDDI. \footnote{Note that, so far, no scheme for gravity gradient mitigation~\cite{DAmico2017Canceling, Roura2017Circumventing, Overstreet2018Effective} has been included for the considered interferometer setup; thus, each constituent AIF will have a non-trivial wave-packet separation at the output port, leading to a possible loss of contrast. However, applying the same frequency shift to both interferometers in the mirror pulse will not result in a different differential phase shift, therefore opening up the possibility to include a (common) detuning for the two constituent AIFs, increasing each one's contrast, without altering the phase shift. } 

\begin{table}
	\begin{tabularx}{\linewidth}{|c|c|c|Y|c|}
	   \hline
	   \multicolumn{5}{|c|}{\textbf{Phase comparison of MZI and SDDI}} \\
	   \hline
	   ~~MZI~~ & ~~SDDI~~ & Phase & Magnitude [rad] & Differential signal \\ 
	   \hline
       $2$ & $2$ & $N k g T_R^2$ & $1.4 \times 10^7$ & $0$ \\  
       \hline 
       $2$ & $2$ & $N k z_0 \Gamma_0 T_R^2$ & $20$ & $0$ \\  
       \hline 
       $2$ & $2$ & ~$N k v_0 \Gamma_0 T_R^3$~& $14$ & $0$ \\
	   \hline
	   $-\frac{7}{6}$ & $-\frac{7}{6}$ & $N k g \Gamma_0 T_R^4$ & $14$ & $0$ \\
	   \hline
       \rowcolor[gray]{0.9}
	   $2$ & 0 & $\frac{N^2 \hbar k^2 \Gamma_0 T_R^3}{m}$ & $1.5 \times 10^{-2}$ & $2$ \\
	   \hline
	   $- 6$ & $- 6$ & $\frac{N \omega_R g^2 T_R^3}{c^2}$ & $2.3 \times 10^{-9}$ & $0$ \\
	   \hline
	   $6$ & $6$ & $\frac{N \omega_R g v_0 T_R^2}{ c^2} $ & $2.4 \times 10^{-9}$  & $0$  \\
	   \hline
	   $10$ & $0$ & $\frac{N^2 \omega_R \hbar k g T_R^2}{m c^2} $ & $1.1 \times 10^{-12}$ & $10$ \\
	   \hline
	   $-4$ & $0$ & $\frac{N^2 \omega_R \hbar k v_0 T_R}{m c^2}$ & $1.1 \times 10^{-12}$ & $-4$ \\
	   \hline
        0 & $4$ & $\frac{N^3 \omega_R \hbar^2 k^2 T_R}{m^2 c^2} $ & $5.7 \times 10^{-16}$ & $-4$ \\
	   \hline
	\end{tabularx}
	\caption{Comparison of phases in the MZI and SDDI in Fig.~\ref{Fig: Tidal_Phase_Geometry}, split into different proportionalities. The first two columns describe the prefactor of phase shift contributions given in the third column, which is present in each AIF phase output. The magnitude denotes the absolute value of the expression in the `Phase' column with assumed numerical values: $N=1$, $\omega_R = 10^7 \, \mathrm{Hz}$, $k = 4 \times 10^6 \, \mathrm{m}^{-1}$, $m = 87 \, \mathrm{amu}$, $T_R = 0.6 \, \mathrm{s}$, $z_0 = 5 \, \mathrm{m}$, $v_0 = 6 \, \mathrm{m/s}$, $g = 9.81 \, \mathrm{m/s^2}$ and $\Gamma_0 = - 2.7 \times 10^{-6} \, \mathrm{Hz}^2 = -2.7 \times 10^3 \, \mathrm{E}$. Gravitational gradients are given in Eötvös ($1 \, \mathrm{E} = 10^{-9} \, \mathrm{Hz}^2$). The last column comprises the prefactor of the phase expression in a differential measurement setup between MZI and SDDI.}
	\label{Table: All Phases Dimensionless Parameters}
\end{table}

The results of a comprehensive account of phases along the lines of~\cite{Werner2024Atom} are summarized in Table~\ref{Table: All Phases Dimensionless Parameters}. The first contribution, $N k g T_R^2$, is the well known phase due to linear gravity. The next three phases connect the gravity gradient with the initial conditions $z_0$, $v_0$ and the linear gravitational acceleration -- each of which are identical for the MZI and SDDI. The fifth contribution is the phase of interest and provides the dominant contribution in the differential signal of both interferometers. The remaining phases, of which some contribute to the differential propagation phase Eq.~\eqref{eq: Tidal_Phase_Ideal_Potential}, are related to the Doppler effect or relativistic corrections and are orders of magnitude smaller than the fifth phase. More details on each of the phase contributions and their origin are given in the Appendix~\ref{Appendix_A}.

The phase in Eq.~\eqref{eq: Tidal_Phase_Ideal_Potential} corresponds to the dominant phase discussed by Asenbaum et al. in~\cite{asenbaum2017phase}, where it was phrased as a `tidal phase'. In this reference, the phase arises from the differential signal of two spatially separated MZIs, one of which is in close proximity to a gravitational test mass inducing a local gravitational gradient~\footnote{The gravity gradient $\Gamma_0$ is denoted by Asenbaum et al.~\cite{asenbaum2017phase} by $T_{zz}$. The factor of $4$ difference in the expression for the tidal phase results from a difference in the number of imprinted photon momenta in the MZIs.}. 
The same phase has also been discussed before by Dimopoulos et al.~\cite{dimopoulos2008general}, where it was referred to as a `1st gradient recoil'. It is worth noting that one can interpret this phase as a `gradient correction' of the recoil phase $\Delta \Phase_\text{recoil} = \hbar N^2 k^2 T_R/m$, which is at the heart of the AIF measurements of the fine-structure constant~\cite{parker2018measurement, morel2020determination}. 

Having an analytic form of this phase shift, see Eq.~\eqref{eq: Tidal_Phase_Ideal_Potential}, reveals the key advantage of this geometry -- namely, the direct proportionality of the phase shift to $\Gamma_0$ and the very accurately known quantities $k$, $T_R$ and $\hbar/m$, enabling to measure the gravity gradient with an accuracy determined by the phase shift measurement. To emphasize this point once more: Typically, phase shifts that include the gravity gradient depend either on the initial conditions $z_0$ and $v_0$, which have a comparably high measurement uncertainty, particularly for smaller baselines \cite{loriani2020colocation}, or they exhibit a non-linear dependence on the gravitational field, i.e. the scale with $g(z) \cdot \Gamma(z)$ as in~\cite{Figure8}, making the inversion complicated for complex gravitational fields. 

\begin{figure}
    \centering
    \includegraphics[width = \linewidth]{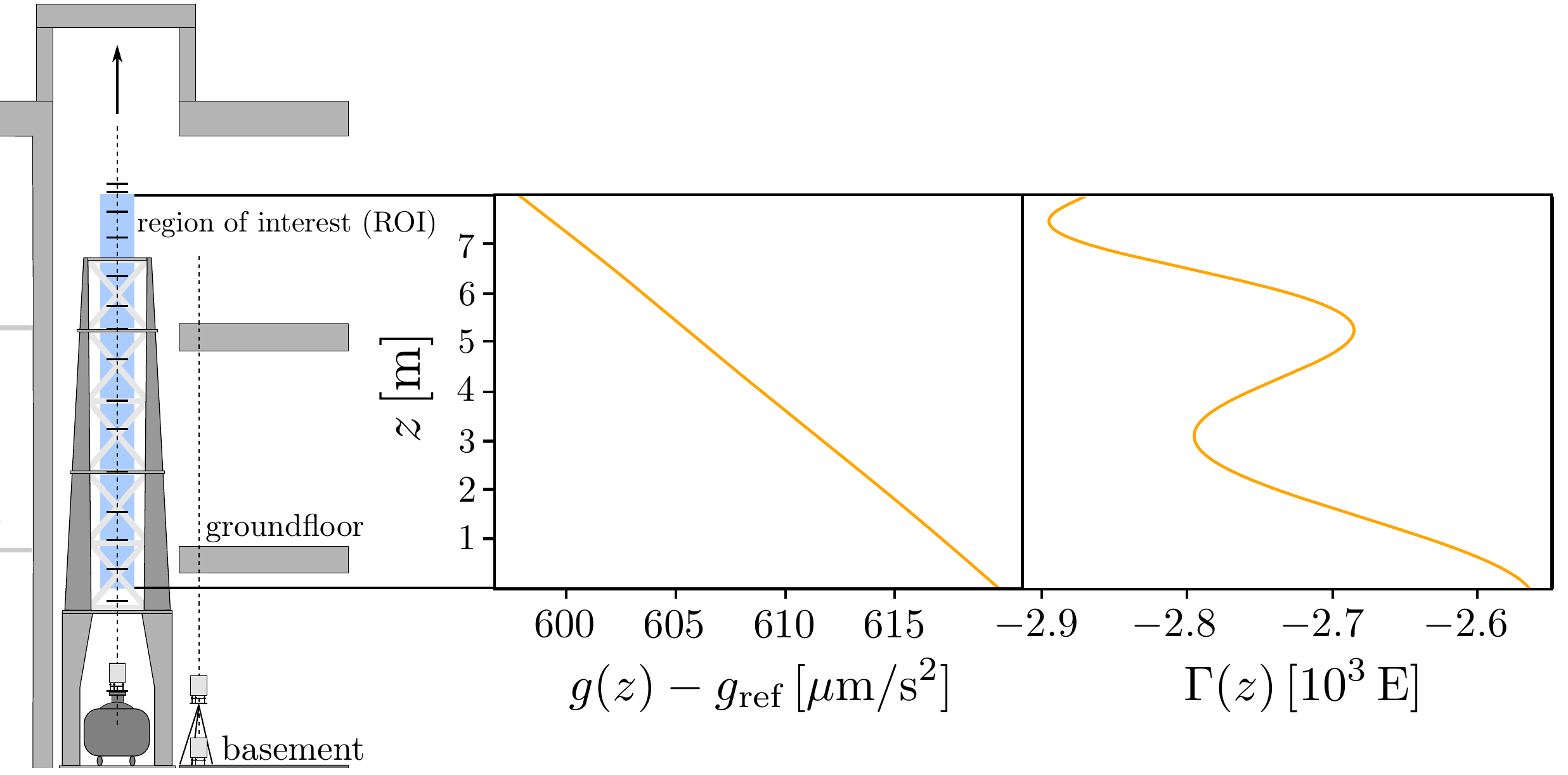}
    \caption{Gravitational acceleration $g(z)$ and gravitational gradient $\Gamma(z)$ as functions of height in the region of interest (ROI) ($0 \, \mathrm{m}$ - $8 \, \mathrm{m}$) of VLBAI with a reference acceleration of $g_\text{ref} = 9.812 \, \mathrm{m/s^2}$. $g(z)$ is interpolated by a polynomial fit. Building cross-section taken from~\cite{schilling2020vertical} and adapted.}
    \label{fig: VLBAI_gravity_model}
\end{figure}

We have now demonstrated how the gravity gradient can be extracted using a novel differential setup, utilizing only the spatial dimensions of a single interferometer. However, up to this point, we have considered an idealized gravitational potential with a constant gravitational gradient. In real-life experiments, this assumption will inevitably be violated to varying extents, as will be discussed in the next section at the example of the VLBAI setup located in Hannover.
The (classically) measured gravitational field in this experimental setup offers a chance to analyze whether the CGI continues to extract information about the gravitational gradient, to understand the averaging process along the atomic trajectory, and, most importantly, to identify any errors that may arise in the interpretation of real-life phase shift data.

\subsection{Gravitational background of the VLBAI Hannover}\label{Section_4}

In VLBAI facilities under construction around the world~\cite{zhan2020zaiga, badurina2020aion, dickerson2013multiaxis}, despite the best efforts to thermally and magnetically shield the atoms from the outside world, gravitational non-linearities can hardly be compensated for by additional masses. Especially temporarily varying mass distributions, such as ground water or even laboratory equipment and concrete structures may alter the gravitational field that the atoms experience during each AIF sequence. At the VLBAI facility in Hannover, a high-precision measurement campaign was carried out with classical sensors to understand the gravitational field -- and its fluctuations -- along the $10$\,m baseline of the interferometer~\cite{schilling2020vertical, lezeik2023understanding}. Fig.~\ref{fig: VLBAI_gravity_model} displays the measured gravitational non-linearity of the gravity gradient as a function of height, which varies in the range of about $10^{-7} \, \mathrm{s}^{-2}$, i.e., $10^{-6} g/\mathrm{m}$. The variations correlate with the building structure, cf. Fig.~\ref{fig: VLBAI_gravity_model}.

In the following we will discuss our differential measurement scheme for the VLBAI Hannover. We start by analyzing multiple CGI sequences from Fig.~\ref{Fig: Tidal_Phase_Geometry}, where the MZI spans height difference of $\Delta h$, with varying initial heights $z_0$. The apex of each atomic trajectory will be obtained at $t = T_R$, which is achieved by setting $v_0 = g T_R$ and $T_R = \sqrt{2 \Delta h / g}$. For simplicity we will always analyze such trajectories in this section, s.t. one can view $\Delta h$ and $T_R$ interchangeably. Due to the non-uniform nature of the gravitational potential, it is not immediately evident how to convert a phase shift measurement into an accurate estimate of the gravitational quantity of interest. This challenge arises because the atoms sample a macroscopic portion of the gravitational field along their path, effectively interacting with non-trivially averaged versions of $g$ and $\Gamma$ throughout their trajectories. Note that the (idealized) dominant phase shift in Eq.~\eqref{eq: Tidal_Phase_Ideal_Potential} is cubic in time, which hints to the fact that the averaging procedure is governed by the time exponent of the idealized phase shift. Based on our numerical simulations, we can confirm that the gravitational gradient at a certain height has a direct correlation with the phase shift of the CGI, where the atomic initial height is shifted by the cubic mean of the atomic position along its trajectory 
\begin{align}
    \norm{z(t)}_3 &= 
    \qty(\frac{1}{2 T_R} \int\limits_0^{2 T_R} \abs{z(t) - z_0}^3 \,  \dd t)^{1/3} \approx 0.77 \, \Delta h. \label{eq: Cubic Norm}
\end{align} 
Connecting this finding with the previously introduced scale factor $f$ one can define an estimator for the gravity gradient by
\begin{align}
    \hat{\Gamma} (z_0) = \frac{\Delta \Phase\qty(z_0 - \norm{z(t)}_3)}{f}.
    \label{eq: Definition Estimator gravity gradient}
\end{align}
Note that $f$ arises from the idealized description of the CGI and should therefore yield an especially good approximation for the actual, but unknown, scale factor for small $\Delta h$. 
\begin{widetext}
\centering
\begin{minipage}[T]{0.8\linewidth}
    \begin{figure}[H]
        \centering
        \includegraphics[width=\linewidth]{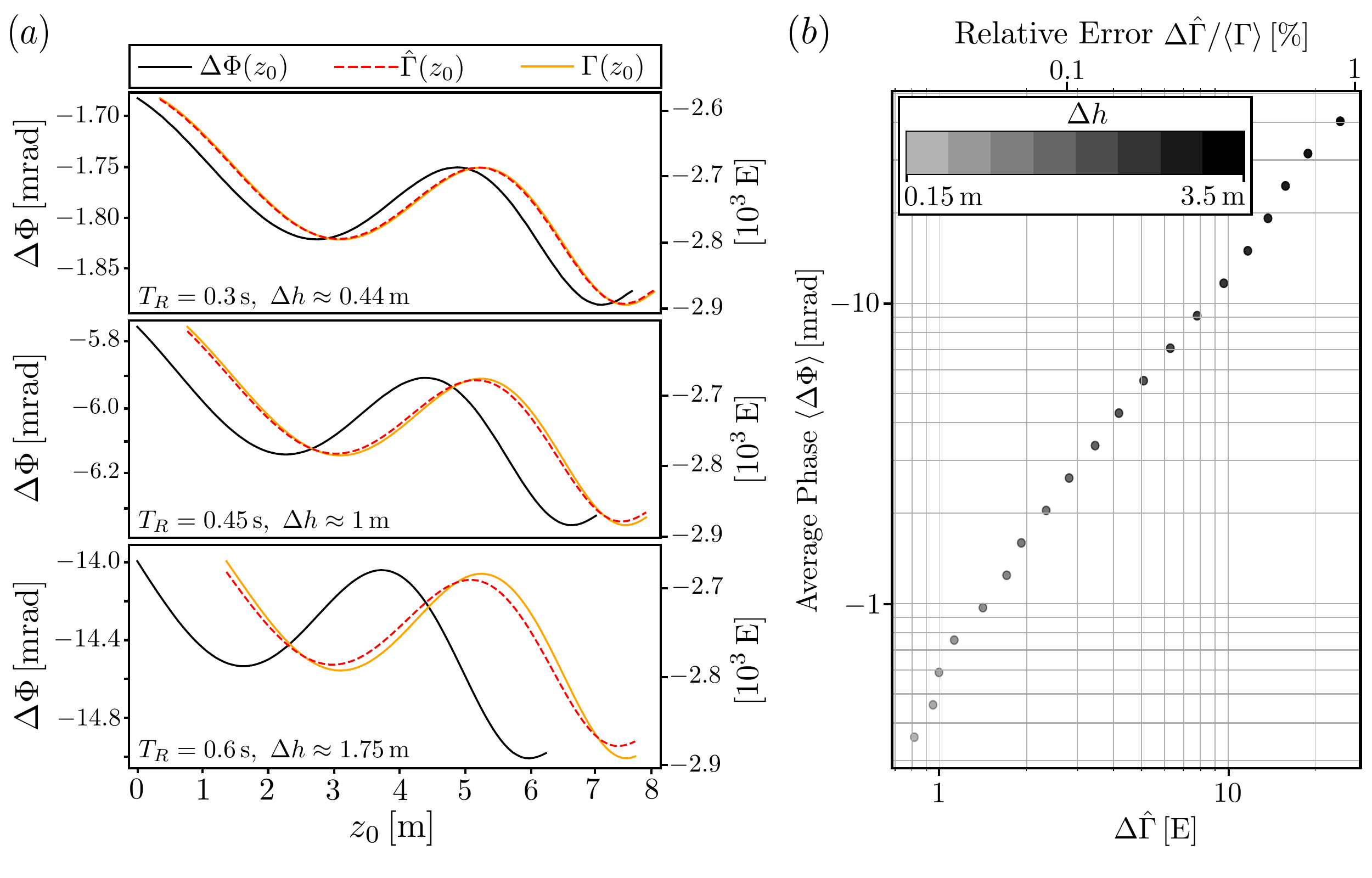}
        \caption{(a) Comparison of the measured phase shift $\Delta \Phase(z_0)$ (black), the gravity gradient $\Gamma(z_0)$ from Fig.~\ref{fig: VLBAI_gravity_model} (orange), and the estimator for the gravity gradient $\hat{\Gamma}(z_0)$ from Eq.~\eqref{eq: Definition Estimator gravity gradient} (red dashed) for three different values of $T_R$ (and therefore $\Delta h$), i.e. different choices of measurement resolution. (b) Phase shift magnitude for CGIs with varying baselines $\Delta h$ and corresponding root mean-square error in the estimation of the gravity gradient. $\Delta \hat{\Gamma}$ is averaged over all possible initial heights in the ROI obtainable with a baseline of $\Delta h$ and $\langle \Gamma \rangle = 2.75 \times 10^3 \, \mathrm{E}$ is the magnitude of the mean gravitational gradient of the facility.}
        \label{fig: Tidal_Phase_Analysis}
    \end{figure}    
\end{minipage}
\end{widetext}
Fig.~\ref{fig: Tidal_Phase_Analysis}~(a) shows the excellent agreement with the correct gravity gradient $\Gamma(z_0)$ for small baselines $\Delta h$, i.e. high spatial resolutions. We furthermore illustrate how increasing the baseline -- and thus the average phase shift -- results in a higher root mean-square error $\Delta \hat{\Gamma}$ in the estimation of the gravity gradient in Fig.~\ref{fig: Tidal_Phase_Analysis}~(b).

These findings highlight the critical importance of having a thorough understanding of the gravitational environment in VLBAI facilities, since the connection between the measured phases and the corresponding height this estimation belongs to were previously unclear. This knowledge is particularly crucial when the objective is to detect signals from additional test masses or even gravitational waves.

\section{Discussion}\label{Section_5}

We have introduced a novel differential setup to measure gravitational curvature and have simulated its behavior in a complex gravitational field. Furthermore, we defined an estimator for the gravity gradient that aligns with the true value within a 1\% margin for interferometers with baselines up to $3.5 \, \mathrm{m}$. 
The question of determining the appropriate averaging procedure to achieve alignment between the AIF phase and the gravitational signal, as in our case with the gravitational gradient case, is inherently complex, especially for arbitrary gravitational fields. Future work on this topic will be necessary, especially if one wants to measure gravitational waves or dark matter with earthbound experiments, with more elaborate interferometer geometries. This is because gravitational perturbations and non-linearities cannot be effectively shielded in those setups and must therefore be characterized with high accuracy.
For example, in the VLBAI facility in Hannover targeting among others high precision gravimetry, the gravity gradient changes by approximately $300 \, \mathrm{E}$ over the whole baseline. Moreover, constructions such as underground tunnels, modify the gravitational field and can be detected by gravity gradient measurements as exemplified by Stray et al.~\cite{Stray2022} employing a conventional gradiometer using two MZIs. (There, deviations of about $150 \, \mathrm{E}$ required a phase resolution of $17.5 \, \mathrm{mrad}$.)

Another alternative strategy to measure the gravity gradient involves the mitigation techniques~\cite{DAmico2017Canceling, Roura2017Circumventing, Overstreet2018Effective}. For the idealized gravitational potential from Eq.~\eqref{eq: Ideal_Gravitational_Potential}, these schemes modify one of the AIF pulse as $k \longmapsto \qty(1 + \Gamma_0 T_R^2/2) \, k$, reducing the wave-packet separation at the output port to achieve higher contrast. By scanning through different pulse detunings and identifying the highest contrast of the interference signal, one can infer the value of $\Gamma_0$ from the optimal detuning frequency. This approach is experimentally simple, but requires multiple AIF experiments to scan various detuning frequencies. The repetition presents a challenge, especially for time-varying gravitational fields, which may cause temporal variations in $\Gamma(z)$. Also, similar to this analysis, one needs to analyze which averaging procedure for $\Gamma(z)$ is involved for the detuning parameter that results in the highest contrast. Additionally one should keep in mind that not only a gravitational gradient would cause a misalignment at the output port, but a variety of different effects, ranging from uncontrolled magnetic field fluctuations to imperfect laser systems, could lead to a wave-packet separation, therefore masking the true value of the gravitational gradient.

Until now, we have assumed the gravitational background near the interferometric baseline to be constant in time. This assumption, however, is not valid, especially for large experimental setups with baselines of 100 meters or more~\cite{mitchell2022magis, abe2021matter}. Variations in ground and surface water $\rho_\text{Water}(t)$, seismic activity $\rho_\text{Earth}(t)$, and even air pressure differentials $\rho_\text{Air}(t)$ can significantly impact the experimental outcomes. It could therefore be beneficial to include an array of these newly described AIFs with an extension of $\Delta h$ and separation $\Delta l$ along the baseline of a large scale experiment. Ideally, this array would be located in a parallel shaft, measuring the gravitational field in real time, while other interferometric experiments are done in the main experimental facility. This array of AIFs should be seen as an integral part of the experimental setup and would be used to gauge and interpret the phase shifts of the other measurements. 

Depending on the frequency of variations in the gravitational potential, $\Delta h$ and $\Delta l$ can be adjusted suitably to obtain a time- and height-resolved measurement of the gravitational field along the baseline. However, the temporal fluctuations of the gravitational field can -- a priori -- span a broad frequency domain. Consider, for the moment, that one wants to resolve changes in the gravitational field with a frequency centered around $\nu$, and that each AIF run takes a time $2 T_R = 2 \sqrt{2 \Delta h / g}$. Firstly, we know that $\nu^{-1} > 2 T_R$, which constrains $T_R$, i.e., $\Delta h$. This can be challenging for very high frequencies $\nu$, as a smaller $T_R$ results in a smaller phase shift, which must still be greater than the measurement uncertainty. Assuming a minimal phase resolution of $1 \, \mathrm{mrad}$, $N = 4$, and the phase output of the AIF being dominantly given by Eq.~\eqref{eq: Tidal_Phase_Ideal_Potential}, this would require a minimal interferometer time of $T_R \geq 0.3 \, \mathrm{s}$, corresponding to a maximum variation frequency of the gravitational field of $\nu \le 3.3 \, \mathrm{Hz}$. This would enable measurements of Earth's primary and secondary micro-seismic frequency peaks, which are both below $1 \, \mathrm{Hz}$, see~\cite{mitchell2022magis}. Note that multiple concurrent interferometer setups like those could, however, improve the sampling frequency and allow for resolutions of even faster gravitational field fluctuations.

Secondly, the choice of $\Delta l$ depends on the complexity of the (static) gravitational potential and the measurement uncertainties of $g(z)$ and $\Gamma(z)$ given by the previously determined value of $\Delta h$. The separation between each interferometer height should be chosen such that it resolves the spatial and temporal changes, possibly by choosing non-uniform separations between each interferometer, i.e., tighter spacing, when the gravitational field is especially non-trivial in space or time.

Extending this concept, one could strategically position gravitational anharmonicities, such as test masses, near the AIF baseline to explore the intricate interplay between quantum mechanics and gravity with greater precision. Phenomena such as the `gravitational Aharonov-Bohm' effect~\cite{overstreet2022observation} and the fundamental interaction between quantum matter and (classical) gravitational fields require a precise interpretation of phase shifts, possibly reaching sub-mrad scales. Therefore, the analysis presented here serves as a crucial preliminary step towards achieving such goals.

To summarize, we introduced a novel AIF geometry designed to exhibit high sensitivity in the measurement of gravitational gradients. In addition, we performed numerical simulations to analyze the behavior of this AIF sequence in the gravitational field of the VLBAI facility in Hannover, Germany. Our results provide new insights into the interpretation of phase shift data in complex gravitational environments. This analysis serves as a case study for VLBAI, highlighting the critical role of accurate gravitational models in state-of-the-art atom interferometry experiments with baselines longer than 10 meters and shows how one would construct an estimator for the gravity gradient in such non-trivial gravitational fields. It should be noted that we idealized the atom-light interaction by assuming instantaneous and lossless processes. We also disregarded Earth's rotation. Actual VLBAI will experience a variety of different error sources, which need to be included in the theoretical description -- especially when it comes to gravitational uncertainties, i.e., resulting from geographical position or a (gravitationally) noisy lab environment in general and are subject to future work.

%%%%%%%%%%%%%%%%%%%%%%%%%%%%
%%%%% ACKNOWLEDGMENTS %%%%%%
%%%%%%%%%%%%%%%%%%%%%%%%%%%%

\section*{Acknowledgments}

We thank Dorothee Tell and Philip Schwartz for insightful discussions. This work was funded by the Deutsche Forschungsgemeinschaft (German Research Foundation) under Germany’s Excellence Strategy (EXC-2123 QuantumFrontiers Grants No. 390837967), through CRC 1227 (DQ-mat) within Projects No. A05, B07, B09 and through the QuantERA 2021 co-funded project No. 499225223 (SQUEIS), and the German Space Agency (DLR) with funds provided by the German Federal Ministry for Economic Affairs and Climate Action (BMWK) due to an enactment of the German Bundestag under Grants No. 50WM2253A (AI-Quadrat). MW, KH and NG acknowledge funding by the AGAPES project - grant No 530096754 within the ANR-DFG 2023 Programme. 

\section*{Author contributions}

M.W., K.H., N.G., D.S., and E.M.R. initiated the research direction, M.W. and K.H. developed the analytical and numerical modelling, M.W. created the figures and drafted the initial manuscript, A.L. and D.S. provided the gravitational model of the VLBAI Hannover, D.S., E.M.R., N.G., and K.H. supervised the project. All authors actively participated in discussing the results and contributed to the final manuscript.

\section*{Competing interests}

The authors declare no competing interests.

\section*{Data Availability}

The numerical and algebraic phase shift calculations that support the findings of this study are available in~\cite{New_Python_Code}.

%%%%%%%%%%%%%%%%%%%%%
%%%%% APPENDIX %%%%%%
%%%%%%%%%%%%%%%%%%%%%

\appendix
\section{Proof of main statement} \label{Appendix_A}

In the following we justify the approximations done in Eq.~\eqref{eq: Phase_Approximation}. We do so by starting with the findings of the previous analysis~\cite[Table II]{Werner2024Atom}, namely the phase shifts of a MZI and a SDDI without FSL effects in the idealized gravitational potential $\Grav_\text{Ideal}$ from Eq.~\eqref{eq: Ideal_Gravitational_Potential}. We will adapt the results of \cite{Werner2024Atom} for the case of this analysis by setting $\Gamma = \Gamma_0$, $k_B = T_B = 0$, i.e. having no intermediate Bloch oscillations and we will impart twice the amount of momentum in the MZI as compared to the SDDI, resulting in Table~\ref{Table: All Phases Dimensionless Parameters}.

One can see how only term $\#$ 9 -- the phase of interest resulting from Earth's gravitational gradient $\Gamma_0$ -- gives a non-trivial phase shift contribution. Terms $\#\,$19, $\#\,$21 and $\#\,$23 are multiple orders of magnitude smaller, because of their proportionality to the recoil frequency $\omega_R$. We explain in the following where this term originates in greater detail. We calculated phase shifts for the idealized and the VLBAI gravitational potential in the manner described in~\cite{Werner2024Atom}.

Let us start by analyzing the AIF from Fig.~\ref{Fig: Tidal_Phase_Geometry} for the case of the idealized gravitational potential from Eq.~\eqref{eq: Ideal_Gravitational_Potential}. The full atomic trajectory of each path of the AIF needs to be defined piecewise as
\begin{align}
z(t) = 
\begin{cases} 
      z_1(t) & t \in [0, T_R) \\
      z_2(t) & t \in [T_R, 2 T_R),
   \end{cases}
\end{align}
where each $z_i(t)$ depends on multiple quantities, i.e. the number $N$ of imparted photon momenta at the start of the trajectory, the wave-vector $k$ of the photon interaction and the initial conditions $z_0$ and $v_0$ as
\begin{multline}
    z(t, N, k, z_0, v_0) = z_0 + \qty(v_0 + \frac{N \hbar k}{m}) t - \frac{1}{2} g t^2 \\
    - \frac{\Gamma_0}{2} \qty(z_0 t^2 + \frac{1}{3} \qty(v_0 + \frac{N \hbar k}{m})t^3 - \frac{1}{12} g t^4).
\end{multline} 
For simplicity we will from now on assume that the atomic wave-packet -- before any laser interactions -- has initial conditions $z_0 = v_0 = 0$. The four atomic trajectories in the time interval $[0, T_R)$, i.e. $z_1(t)$, will then be given by
\begin{align*}
    z_\text{up}^\text{MZI} \qty(t, 2N, k, 0, 0) &= \frac{2 N \hbar k}{m} t - \frac{g}{2} t^2 - \frac{\Gamma_0}{2} \qty(\frac{2N \hbar k}{3m}t^3 - \frac{g}{12} t^4), \\
    z_\text{low}^\text{MZI} \qty(t, 0, k, 0, 0) &= - \frac{g}{2} t^2 + \frac{\Gamma_0}{2} \frac{g}{12} t^4, \\
    z_\text{up}^\text{SDDI} \qty(t, N, k, 0, 0) &= \frac{N \hbar k}{m} t - \frac{g}{2} t^2 - \frac{\Gamma_0}{2} \qty(\frac{N \hbar k}{3m} t^3 - \frac{g}{12} t^4), \\
    z_\text{low}^\text{SDDI} \qty(t, -N, k, 0, 0) &= - \frac{N \hbar k}{m} t - \frac{g}{2} t^2 + \frac{\Gamma_0}{2} \qty(\frac{N \hbar k}{3m}t^3 + \frac{g}{12} t^4).
\end{align*}
We then define shorthand notations for the heights and velocities of these trajectories at $t= T_R$ via
\begin{align*}
    z^\text{MZI}_\text{up}&= z_\text{up}^\text{MZI} \qty(T_R, 2N, k, 0, 0),  &
    v^\text{MZI}_\text{up}&= \dot{z}_\text{up}^\text{MZI}\qty(T_R, 2N, k, 0, 0), \\
    z^\text{MZI}_\text{low} &= z_\text{low}^\text{MZI} \qty(T_R, 0, k, 0, 0), &
    v^\text{MZI}_\text{low}&= \dot{z}_\text{low}^\text{MZI}\qty(T_R, 0, k, 0, 0), \\
    z^\text{SDDI}_\text{up} &= z_\text{up}^\text{SDDI} \qty(T_R, N, k, 0, 0), &
    v^\text{SDDI}_\text{up}&= \dot{z}_\text{up}^\text{SDDI}\qty(T_R, N, k, 0, 0), \\
    z^\text{SDDI}_\text{low} &= z_\text{low}^\text{SDDI} \qty(T_R, -N, k, 0, 0), & 
    v^\text{SDDI}_\text{low}&= \dot{z}_\text{low}^\text{SDDI}\qty(T_R, -N, k, 0, 0).
\end{align*}

Having those quantities one is able to write down the atomic trajectories in the time interval $[T_R, 2 T_R)$, i.e. $z_2(t)$, as
\begin{subequations}
\begin{align}
    z_\text{up}^\text{MZI} \qty(t) &= z\qty(t - T_R, -2N, k, z^\text{MZI}_\text{up}, v^\text{MZI}_\text{up}) \\
    z_\text{low}^\text{MZI} \qty(t) &= z\qty(t - T_R, 2N, k, z^\text{MZI}_\text{low}, v^\text{MZI}_\text{low}) \\
    z_\text{up}^\text{SDDI} \qty(t) &= z\qty(t - T_R, -2N, k, z^\text{SDDI}_\text{up}, v^\text{SDDI}_\text{up})  \\
    z_\text{low}^\text{SDDI} \qty(t) &= z\qty(t - T_R, 2N, k, z^\text{SDDI}_\text{low}, v^\text{SDDI}_\text{low}),
\end{align}
\end{subequations}
One is now ready to calculate the phase shifts of each interferometer, assuming two-photon Bragg transitions, instantaneous laser interactions and infinite light speed. When we display phase shifts, as calculated from~\cite{Werner2023Dataset}, we will set $v_0 \neq 0$ and only show phases which consist of three or less dimensionless parameters, as defined in~\cite{Werner2024Atom}.

\vspace{0.5cm}

\subsection*{Separation phase}

We start with the separation phase of each IF, which one can calculate via
\begin{align}
    \Delta \Phase_\text{Sep} = \frac{m}{\hbar} \Delta z \, v_\text{aver},
\end{align}
where $\Delta z$ is the separation at the output port and $v_\text{aver}$ is the average output velocity of the two output ports. For the MZI and SDDI we obtain the separation phases of
\begin{subequations}
\begin{align}
	\Delta \Phase^\text{MZI}_\text{Sep} &= - 2 \Gamma_0 v_0 N k T_R^3 + 4 g N k \Gamma_0 T_R^4 + 8 \frac{N \omega_R g^2 T_R^3}{c^2} \nonumber \\
	& \quad - 4 \frac{N \omega_R g v_0 T_R^2}{c^2} 
    + \qty(4 v_0 - 8 g T_R) \frac{N^2 \omega_R \hbar k T_R}{m c^2}, \\
	\Delta \Phase^\text{SDDI}_\text{Sep} &= - 2 \Gamma_0 v_0 N k T_R^3 + 4 g N k \Gamma_0 T_R^4 + 8 \frac{N \omega_R g^2 T_R^3}{c^2}\nonumber \\
    &\quad - 4 \frac{N \omega_R g v_0 T_R^2}{c^2},
\end{align}
\end{subequations}
\begin{widetext}
\begin{minipage}[T]{0.95\linewidth}
\begin{figure}[H]
    \centering
    \includegraphics[width=1\linewidth]{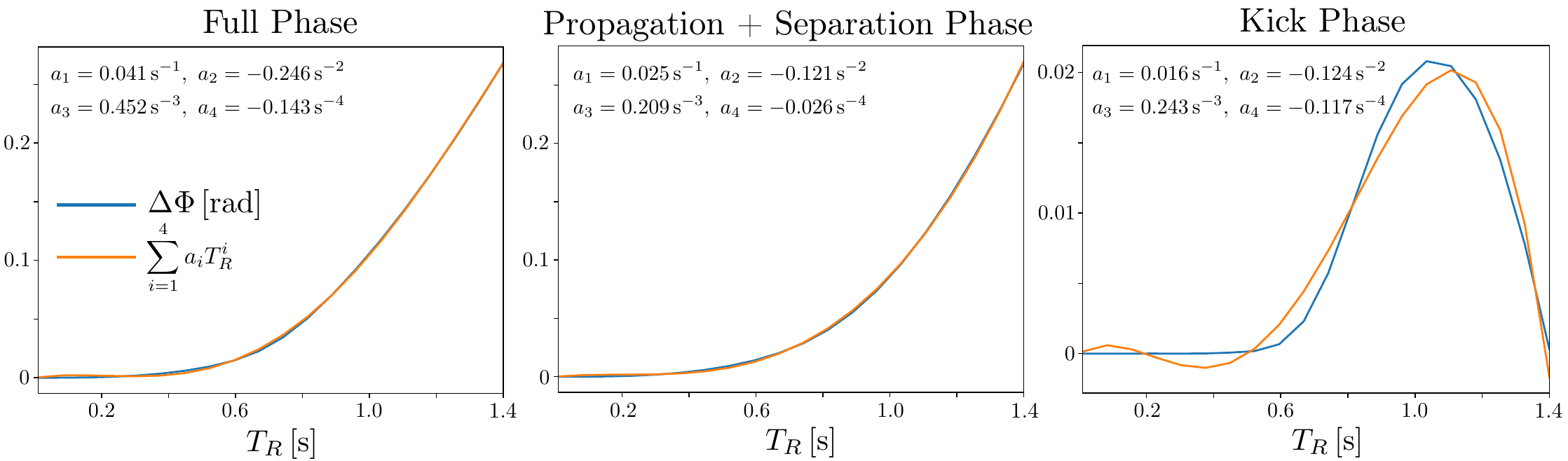}
    \caption{Phase shift simulation (blue) of the CGI in the gravitational field of the VLBAI from Fig.~\ref{fig: VLBAI_gravity_model} as a function of $T_R$ for fixed $N=1$, $z_0=0$ and $v_0 = 13.8 \, \mathrm{m/s}$. A polynomial of fourth order (orange) is fitted to the phase shift. One can see how the phase shift is dominated by propagation and separation phase, whereas the kick phase contributes of at most one order of magnitude below.}
    \label{Fig: Phase_Contributions}
\end{figure}   
\end{minipage}
\end{widetext}
resulting in a relative phase shift
\begin{align}
	\Delta \Phase^\text{MZI}_\text{Sep} - \Delta \Phase^\text{SDDI}_\text{Sep} = \qty(4 v_0 - 8 g T_R) \frac{N^2 \omega_R \hbar k T_R}{m c^2}.
\end{align}
One also sees how the gravity gradient effects both separation phases in completely the same manner. The separation phase itself will, however, in modern experiments always be mitigated, since a non-negligible separation on the output port would lead to a substantial loss of contrast.

Note that it was shown in~\cite{dimopoulos2008general} how the separation phase can be viewed as the `missing part' of the closed propagation phase integral. Whenever we have a substantial separation at the output port it will therefore make sense to analyse the sum of the propagation and separation phase, since they arise from the same intrinsic mechanism.

\subsection*{Kick phase}

The Kick phase can be calculated by the weighted sum of light field phases $\pm \Phi(t_\text{int}, z_\text{int})$ at each interaction time $t_\text{int}$ and height $z_\text{int}$, counted positively if a photon is absorbed in the process and negatively if a photon is emitted. Note that the frequencies of each laser pulse need to be Doppler-corrected, s.t. the desired momentum transfer is achieved resonantly. 
Calculating this for the MZI and SDDI yields
\begin{subequations}
\begin{align}
	\Delta \Phase^\text{MZI}_\text{Kick} &= 2 g N k T_R^2 + 2 N k z_0 \Gamma_0 T_R^2 + 2 \Gamma_0 v_0 N k T_R^3 \nonumber \\ 
    &\quad - \frac{7}{6} g N k \Gamma_0 T_R^4 - 6 \frac{N \omega_R g^2 T_R^3}{c^2} + 6 \frac{N \omega_R g v_0 T_R^2}{c^2} \nonumber \\
	&\quad + 6 \frac{N^2 \omega_R \hbar k g T_R^2}{m c^2} - 4 \frac{N^2 \omega_R \hbar k v_0 T_R}{m c^2}, \\
	\Delta \Phase^\text{SDDI}_\text{Kick} &= 2 g N k T_R^2 + 2 N k z_0 \Gamma_0 T_R^2 + 2 \Gamma_0 v_0 N k T_R^3 \nonumber \\
    &\quad - \frac{7}{6} g N k \Gamma_0 T_R^4 - 6 \frac{N \omega_R g^2 T_R^3}{c^2} + 6 \frac{N \omega_R g v_0 T_R^2}{c^2}  .
\end{align}
\end{subequations}
The differential phase shift is therefore given by
\begin{align}
    \Delta \Phase^\text{MZI}_\text{Kick} - \Delta \Phase^\text{SDDI}_\text{Kick} 
    = \frac{\hbar k \omega_R T_R}{m c^2} \qty(6 g T_R - 4 v_0)
\end{align} 
and does not depend on $\Gamma_0$ at all. Note that these are the phases $\#$ 19, $\#$ 21 and $\#$ 23 from Table~\ref{Table: All Phases Dimensionless Parameters} and are orders of magnitude smaller than the phase of interest.

\subsection*{Propagation phase}

The most interesting phase shift contribution is the propagation phase which as already introduced in Eq.~\eqref{eq: Definition_Propagation_phase}. We will now show in more detail that all phases including linear gravitational acceleration drop out, i.e. especially show Eq.~\eqref{eq: Propagation_phase_difference_statement}. The propagation phases evaluate to
\begin{subequations}
\begin{align}
    \Delta \Phase_\text{Prop}^\text{SDDI} &= -4 g \Gamma_0 N k T_R^4 + 2 N k v_0 \Gamma_0 T_R^3 - 8 \frac{N_R \omega_R g^2 T_R^3}{c^2}  \nonumber \\
	& \quad + 4 \frac{N_R \omega_R g v_0 T_R^2}{ c^2}+ 12 \frac{N_R^2 \omega_R \hbar k_R g T_R^2}{m c^2} \nonumber \\
    &\quad - 4 \frac{N_R^2 \omega_R \hbar k_R v_0 T_R}{m c^2} - \frac{2 \Gamma N_R^2 \hbar k_R^2 T_R^3}{m}, \\
    \Delta \Phase_\text{Prop}^\text{MZI} &= 2 \frac{\hbar N^2 k^2 \Gamma_0 T_R^3}{m} - 4 g \Gamma_0 N k T_R^4 + 2 N k v_0 \Gamma_0 T_R^3 \nonumber \\
    &\quad - 8 \frac{N_R \omega_R g^2 T_R^3}{c^2} + 4 \frac{N_R \omega_R g v_0 T_R^2}{ c^2} \nonumber \\
	& \quad + 4 \frac{N_R^3 \omega_R \hbar^2 k_R^2 T_R}{m^2 c^2},
\end{align}  
\end{subequations}
which results in a differential signal equal to the discussed phase and negligible Doppler terms, i.e.,
\begin{multline}
    \Delta \Phase_\text{Prop}^\text{MZI} - \Delta \Phase_\text{Prop}^\text{SDDI} = 2 \frac{\hbar N^2 k^2 \Gamma_0 T_R^3}{m} \\
    + \frac{N^2 \omega_R \hbar k T_R}{m c^2} \qty(12 g T_R - 4 v_0 - 4 \frac{N \hbar k}{m}).
\end{multline}
Note that the this phase was present in multiple prior investigations of the MZI~\cite{dimopoulos2007PRL, hogan2008light}, the interesting aspect is rather, that the symmetric momentum imprint of the SDDI nullifies this phase contribution. We therefore reproduced all terms in the Table~\ref{Table: All Phases Dimensionless Parameters} and showed their respective origin.

\subsection*{Transition to non-ideal gravitational potentials}

One has previously seen how the dominant differential phase arises from the propagation integral along the beyond first order potential terms of the Lagrangian and the macroscopic height difference of the MZI at the mirror pulse of $\sim 2N \hbar k T_R/m$. The propagation phase for a non-trivial gravitational potential will therefore always take the form 
\begin{align*}
    \Delta \Phase_\text{Prop}^\text{MZI} \approx \frac{m\Grav^{(2)}}{2 \hbar} \qty( \frac{2N \hbar k T_R}{m} )^2 +\frac{m\Grav^{(3)}}{6 \hbar} \qty( \frac{2N \hbar k T_R}{m} )^3 + \dots,
\end{align*}
whereas $\Delta \Phase_\text{Prop}^\text{SDDI}$ lacks those kind of terms. The kick phase, however, only depends on the atomic positions and both paths of the AIF are affected by gravity in a similar way because of the small spatial separation between the two arms on the scale of $2N\hbar k T/m$. Note that the expression for the phase in Eq.~\eqref{eq: Tidal Phase Abstract Definition} is essentially given by the propagation phase, where the kinetic and the linear gravitational part of the Lagrangian drop out, as we have seen for the case of the idealized potential. 

We simulate the CGI in the gravitational field of the VLBAI from Fig.~\ref{fig: VLBAI_gravity_model}. The numerics are done in Python~\cite{New_Python_Code} and use a time discretization of the interval $[0, 2 T_R]$ into a certain number of sub-intervals. For the VLBAI gravity profile we have seen that a number of 20.000 time steps is sufficient for convergence. Fig.~\ref{Fig: Phase_Contributions} displays how the full phase output -- and its constituent phase shifts -- scale as a function of $T_R$. FSL phases are omitted in this simulation, i.e., all atom-light interactions happened at time instances: $t=0, T_R, 2 T_R$.

The approximation in Eq.~\eqref{eq: Phase_Approximation} fails in a few cases. Firstly -- a rather special case -- if the separation phase becomes too large, it will dominate the propagation integral from Eq.~\eqref{eq: Phase_Approximation}. This case is, however, experimentally trivial, since a substantial separation phase would yield zero contrast anyway. Secondly, if the gravitational potential is (artificially) designed such that it accelerates one of the interferometer arms very differently from the other one. This would give non-trivial results in the propagation phase, as well as in the kick phase. Of course, this effect is also quite artificial, since the atomic paths need to interfere at the output port nevertheless.

\section{FSL phase and mitigation scheme}\label{Appendix_B}

This analysis assumed infinitely fast laser beams and completely simultaneous interactions at the times $t = 0, T_R, 2 T_R$ for each IF path. This is, however, not the case because of the finite speed of light. Deviations from the `perfect' interaction times will ultimately lead to additional phase shifts, known as FSL phases~\cite{tan2016finite}. The FSL phases depend on the experimental conditions, photon path lengths, mirror positions and, most importantly, on the type of laser interaction used to drive the beam splitting and mirroring devices. 

\begin{figure}
    \centering
    \includegraphics[width=\linewidth]{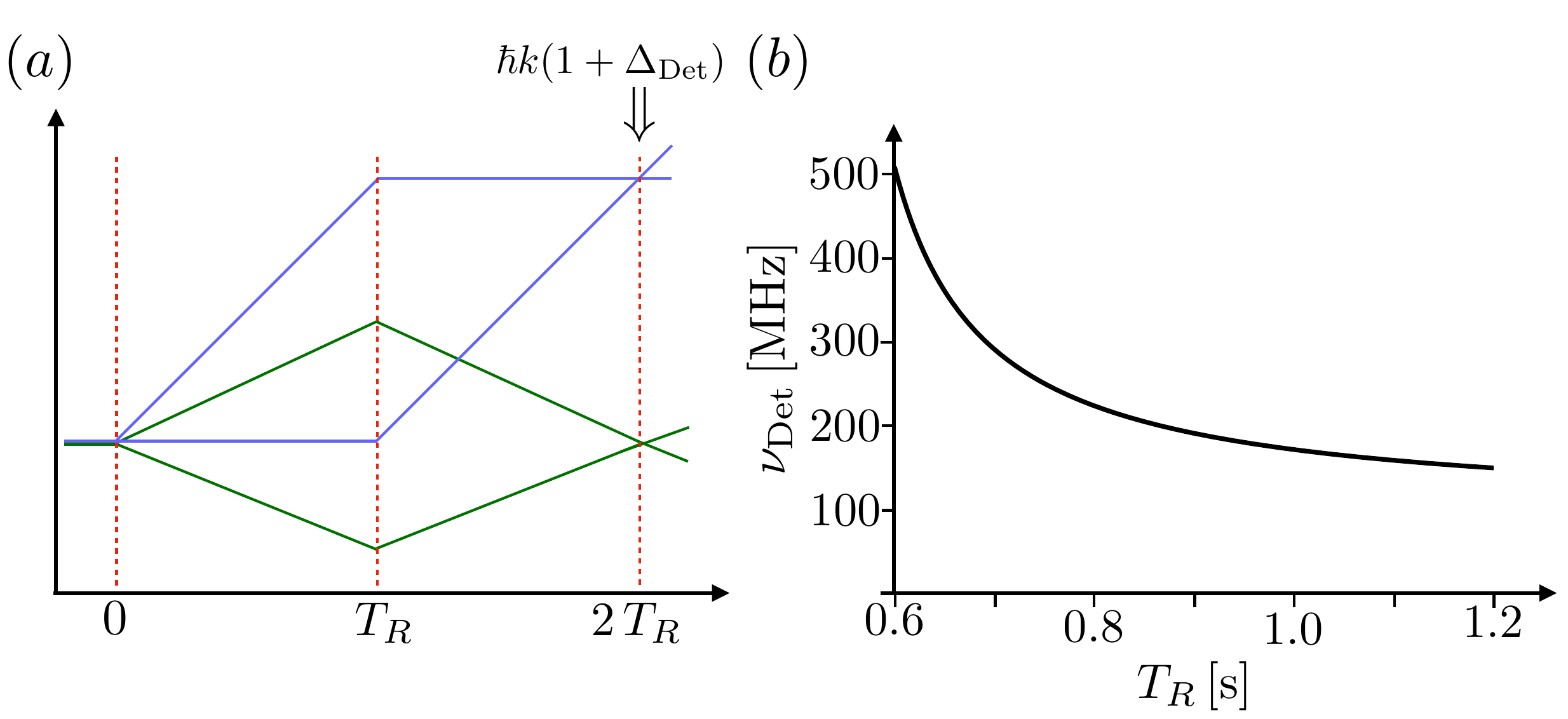}
    \caption{(a) Schematic depiction of the mitigation scheme using a detuning of the last IF pulse. (b) Optimal detuning frequency for a single interaction ($N = 1$) $\nu_\text{Det}\qty(v_0, T_R) = c k \DetParam \qty(v_0, T_R)$ as a function of $T_R$ for fixed $v_0 = 5 \, \mathrm{m}/\mathrm{s}$.}
    \label{Fig: FSL_Figure}
\end{figure}

\subsection*{Two-photon Bragg transitions}

For the case of two-photon Bragg transitions, i.e. two counter-propagating laser fields with wave vectors $k_1$ and $-k_2$ and an effective wave vector $k = k_1 + k_2$, this phase shift evaluates to
\begin{align}
    \Delta \Phase_\text{FSL} = \frac{4 \hbar N^2 k^2 T_R}{m c} \qty(4 g T_R - v_0 - \frac{N \hbar k}{m}) + \Delta \Phase_0,
    \label{eq: Definition_FSL_Phase}
\end{align}
with an additional, time independent, part 
\begin{align}
  \Delta \Phase_0 = \frac{2 \hbar N^2 k^2}{m c} \qty(2 z_\text{L} - z_0 - z_\text{U}). 
\end{align}
Here we assumed both lasers to be placed at a height $z_\text{U}$ and a retro-reflective mirror at $z_\text{L}$. The time independent part is, per se, not a problem, since it is an invariant quantity for each IF and can be used to accurately calculate the desired quantities from the output phase shift, if its magnitude is known. However, the $T_R$ dependence of the first part can be problematic, since $T_R$, as well as $v_0$, will be variable parameters in the experiment.

These phases result dominantly from the temporal part of the laser phase imprint, i.e. the $\omega_i \Delta t$ parts of the phase of each light field, where $\Delta t$ is the photonic flight time and $\omega_i = c k_i$ is the frequency of each individual light field. 

\subsection*{Mitigation of FSL terms}

Suppose that if one modifies the last IF laser pulse via $\hbar k \longmapsto \qty(1 + \DetParam) \hbar k$, with a dimensionless detuning parameter $\DetParam \ll 1$, one generates an additional (dominant) phase shift at the output port of
\begin{align}
    \Delta \Phase_\text{Additional} = 2 N k T_R \DetParam \qty(v_0 + \frac{N \hbar k}{m} - g T_R).
    \label{eq: Additional_Phase_Shift_Detuning}
\end{align}
It is important to highlight that this detuning parameter must be kept small to ensure a substantial overlap at the final output ports of each constituent IF.
Note that the term in the brackets is usually small, since this term, set to zero, is a typical constraint equation of $T_R$ and $v_0$ for the optimal motion of a launch-mode IF. This is the case, since the apex of the atomic trajectory optimally appears after half of the IF time, i.e. at $T_R$. Having an initial velocity of $v_0 + \frac{N \hbar k}{m}$ then exactly results in the equation 
\begin{align}
    v_0 + \frac{N \hbar k}{m} - g T_R = 0.
    \label{eq: Mitigation_Constraint}
\end{align}
Applying this mitigation strategy one needs to ensure that $v_0$ and $T_R$ are chosen such that Eq.~\eqref{eq: Mitigation_Constraint} is not fulfilled.

The described detuning allows for the calculation of the appropriate form of $\DetParam$ such that the additional phase cancels out the time dependent part of $\Delta \Phase_\text{FSL}$. This cancellation appears if we choose
\begin{align}
     \DetParam \qty(v_0, T_R) = 2 \frac{v_0 + \frac{\hbar k}{m} - 4 g T_R}{v_0 + \frac{\hbar k}{m} - g T_R} \frac{\hbar k}{mc}.
     \label{eq: Optimal_Detuning}
\end{align}
In Fig.~\ref{Fig: FSL_Figure} (b) one can see the optimal detuning frequency $\nu_\text{Det} = c k \DetParam$ of the last IF pulse as a function of $T_R$ for a given initial velocity and number of imprinted photon momenta. One can see how the usual detuning frequency is roughly at the order of hundred MHz.
Note that the optimal detuning becomes infinite for $T_R \approx 0.5$, since Eq.~\eqref{eq: Mitigation_Constraint} would be satisfied given the assumed value of $v_0 = 5  \, \mathrm{m}/\mathrm{s}$. Note that one could, however, detune each pulse via it own small detuning and choose all three variables accordingly. 

Another way to mitigate FSL effects is to change the photonic path lengths in a suitable fashion in order to create additional FSL phase shifts that -- optimally -- cancel the unwanted phase shifts. In the calculation for the different FSL effects we always assumed all laser sources to be on the same, fixed, height above $z_\text{U}$ the IF baseline. Altering this laser height for certain pulses, or mixing two-photon and single-photon transitions could help in mitigating the FSL effects but depends on the on-site possibilities and variability in the experimental setup.

\section{Experimental realisation of the initial beam splitter}\label{Appendix_C}

An experimental implementation of the first beam splitter of the CGI, as depicted in Fig.~\ref{Fig: Tidal_Phase_Geometry}, could involve preparing an atomic cloud containing a mixture of two different hyperfine states. This approach would allow for the individual addressing of both atomic subsystems in the MZI and SDDI geometries. However, it may also introduce the risk of unwanted effects due to magnetic field fluctuations.

One could also perform a short Mach-Zehnder geometry right before the first beam splitter and use the output ports of the MZI as Doppler selective inputs of the new interferometer geometry, i.e., perform the operations
\begin{align}
    \ket{0} \longmapsto \frac{1}{\sqrt{2}} \qty( \ket{0} + \ket{-N} ), \quad 
    \ket{N} \longmapsto \frac{1}{\sqrt{2}} \qty( \ket{N} + \ket{2N} ) ,
\end{align}
in a composite pulse as shown in Fig.~\ref{Fig: Experimental realisation from initial MZI}.
The first setup has the advantage of ideal `colocation' control~\cite{loriani2020colocation}, since the atoms in each AIF will have very narrow uncertainties in initial conditions, because they are initialized in a common trap. The latter setup, however, has the advantage of very good coherence between the two input states.

Given that Bragg scattering is inherently a multi-port process~\cite{siemss2020analytic}, composite pulses could produce an effective four-way beam splitter as needed here. Such generalized beam splitters are created in the laboratory following analytical treatments~\cite{siemss2020analytic} as in Ref.~\cite{pfeiffer2024dichroicmirrorpulsesoptimized} or using optimal quantum control methods~\cite{Saywell2023b,Louie2023a}.

\begin{figure}[H]
    \begin{tikzpicture}[scale=0.6]
		\draw[->, thick] (0, 0) -- (11.2, 0) node[right] {$t$}; 	      % x axis
		\draw[->, thick] (0, 0) -- (0, 8.5) node[left] {$z$};          % y axis

        \draw[color=red, dashed] (1, 0) -- (1, 8);    % Bragg Lasers
        \draw[color=red, dashed] (3, 0) -- (3, 8);    % Bragg Lasers
		\draw[color=purple, dashed, thick] (5, 0) -- (5, 8);	  % Bragg Lasers
		\draw[color=red, dashed] (10, 0) -- (10, 8);  % Bragg Lasers

        \draw[color=black, thick] (0.5, 2.2) -- (1, 2.2) -- (3, 3) -- (5, 3);	   % IF Path MZI Initial Upper
        \draw[color=black, thick] (1, 2.2) -- (3, 2.2) -- (5, 3);	   % IF Path MZI Initial Lower
  
		\draw[color=blue, thick] (5, 3) -- (10, 3);	 
		\draw[color=blue, thick] (5, 3) -- (10, 7);	 
		\draw[color=mygreen, thick] (5, 3) -- (10, 1);  
		\draw[color=mygreen, thick] (5, 3) -- (10, 5);   

        \draw[color=black] (5, -0.5) node {$0$};
        \draw[color=black] (10, -0.5) node {$T_R$};

        \draw[color=black] (10.7, 1) node {$\ket{-N}$};
        \draw[color=black] (10.5, 3) node {$\ket{0}$};
        \draw[color=black] (10.55, 5) node {$\ket{N}$};
        \draw[color=black] (10.7, 7) node {$\ket{2N}$};
    \end{tikzpicture}
    \caption{Schematic depiction of an experimental realisation of the first beam splitter pulse (purple), which is build upon an initial MZI and uses its output ports as inputs of the desired interferometer. Usual beam splitter pulses are depicted in red.}
    \label{Fig: Experimental realisation from initial MZI}
\end{figure}

%%%%%%%%%%%%%%%%%%%%%%%%%
%%%%% BIBLIOGRAPHY %%%%%%
%%%%%%%%%%%%%%%%%%%%%%%%%

\bibliographystyle{apsrev4-1_costum}
\bibliography{bibliography}

\end{document}